\documentstyle[epsf]{elsart}
\input{epsf}

\begin{document}
\begin{frontmatter}

\title{A hydrodynamic approach to the Bose-Glass transition}

\author{Panayotis Benetatos\thanksref{GRANT}},
\author{M. Cristina Marchetti\thanksref{GRANT}}
\address{Physics Department, Syracuse University, Syracuse, NY 13244}


\begin{abstract}
{\it Nonlinear} hydrodynamics is used to evaluate disorder-induced 
corrections to the vortex liquid tilt modulus for {\it finite} screening length
and arbitrary disorder geometry. Explicit results for aligned columnar defects yield a criterion for locating the Bose glass transition line at 
all fields.
\end{abstract}
\thanks[GRANT]{Partially supported by the National Science Foundation through grants 
No. DMR97-30678 and DMR98-05818.}

\end{frontmatter}

The vortex phase diagram of high temperature superconductors shows a rich diversity of phases,
including vortex lattices, liquids and glasses.
Novel types of glasses are also possible because of pinning in 
disordered samples (for a review, see Refs. \cite{CN97,BLATTER94,CRABTREE98}).
Much progress in the theoretical understanding of vortex behavior at low fields has been made by
employing the formal analogy of the statistical mechanics of $(2+1)$-dimensional directed lines 
with the quantum mechanics of $2d$ bosons.\cite{DRN88}
In this mapping, the vortex 
lines traversing the sample along the direction $z$ of the external field, 
${\bf H}_0={\bf \hat{z}} H_0$,
correspond to the imaginary-time world
lines of the quantum particles.\cite{DRN88} 
The thickness, $L,$ of the superconducting
sample is  the inverse temperature, $\beta_B\hbar$, of the bosons,  the vortex line
tension, $\tilde{\epsilon}_1$, represents the boson mass, $m$, and thermal fluctuations in the vortex state
$\propto k_BT$
map onto quantum fluctuations $\propto\hbar$ in the boson system. 
%The entangled flux-line liquid
%corresponds to a boson superfluid, while the disentangled liquid represents the normal fluid.
This ``boson mapping'' has been particularly useful for understanding  the properties
of vortices in superconducting samples with aligned damage tracks from heavy-ion irradiation.
This type of disorder yields a low-temperature Bose-glass phase 
where  every vortex is trapped on a columnar 
defect \cite{Fisher,NV} and vortex pinning is strongly enhanced.\cite{Liou} 
The transition at $T_{BG}$ 
from the entangled vortex liquid
to the Bose-glass is continuous and is signalled by the vanishing of the linear resistivity
and of the inverse tilt modulus, $c^{-1}_{44}$. 
The drawback of the boson mapping in the form used by Nelson and coworkers \cite{DRN88,NV,TN} is that intervortex 
interactions are assumed to be strictly {\em local} in $z$, the magnetic field direction. 
This restricts the application of the results to low fields. It also renders problematic 
the evaluation of the tilt modulus $c_{44}$, which measures the linear response to a 
transverse tilting field ${\bf H}_\perp$ normal to ${\bf H}_0$.
The long-wavelength tilt modulus of the vortex liquid can be written as
$c_{44}=c_{44}^v+B^2/(4\pi)$, where $c_{44}^v=n_0\tilde{\epsilon}_1$, with $n_0=B/\phi_0$ 
the vortex density,
is the single vortex part and the second term represents a compressive contribution. 
By assuming local interactions along $z$, the compressive part of $c_{44}$, which 
dominates at high fields,  is
neglected entirely.  With this approximation, the mathematical analogy 
between vortices and
$2d$ bosons can be exploited further to show that the inverse tilt modulus maps onto
the superfluid density, $n_s$, of the bosons,
with
\begin{equation}
\label{c44v}
{n_0^2\over c_{44}^v}={n_s\over m}.
\end{equation}
The transition as $T\rightarrow T_{BG}^+$ from the entangled vortex liquid to the Bose glass
corresponds then to the transition from a boson superfluid to a localized normal phase of bosons. T\"auber and Nelson have evaluated perturbatively the reduction of $n_s$ due to various types of disorder.\cite{TN} They also showed that for aligned columnar disorder such a perturbative
calculation yields a useful criterion for locating the transition line $T_{BG}(B)$ {\it at
low fields}. 

Larkin and Vinokur \cite{LV} argued that a generalization of Eq. (\ref{c44v}) that incorporates 
the compressive part of $c_{44}$ can be obtained by using the {\it nonlocal}
mapping on vortex lines onto $2d$ bosons introduced some time ago by 
Feigel'man and collaborators.\cite{feigel} These authors showed that the fully nonlocal London 
model of interacting vortex lines can be mapped to a system of $2d$ {\em charged} bosons 
coupled to a massive photon field. The duality between vortices and bosons translates
into
\begin{equation}
\label{c44LV}
c_{44}={B^2\over 4\pi}+{n_0^2\tilde{\epsilon}_1\over n_s},
\end{equation}
where $n_s$ is defined here in terms of the polarization function of the fictitious gauge field. Near a Bose glass transition where $n_s\rightarrow 0$ the second term of 
Eq. (\ref{c44LV}) dominates and results in the divergence
of the tilt modulus. Feigel'man et al. carried out a perturbative calculation
of the reduction of $n_s$ from intervortex interactions in a clean material
(in the limit $\lambda\rightarrow\infty$) \cite{feigel},
but the calculation of the reduction of $n_s$ from disorder in the context
of the charged boson model is cumbersome and does not provide much physical insight.

An alternative approach for modeling interacting vortex arrays is hydrodynamics,
which has proved very useful to describe the long wavelength properties of 
flux-line liquids at high fields.\cite{MCM} Hydrodynamics provides a physically transparent formulation that naturally incorporates the nonlocality of the intervortex interaction. 
The goal of this note is to show how hydrodynamics can be used in a transparent way
to evaluate disorder-induced
corrections to the wave-vector dependent tilt modulus for finite values of the screening length
$\lambda$ and arbitrary disorder geometry. The connection of the hydrodynamic formulation 
to the charged boson formalism will also be discussed. Explicit results are presented
for aligned columnar defects.

The hydrodynamic free energy of the flux-line liquid is a functional of two 
coarse-grained fields, the local areal density of vortices, $n({\bf r})$, and the tilt field,
${\bf t}({\bf r})$, which measures the local deviation of a volume of flux liquid 
from the direction
of the external field, ${\bf H}_0$.\cite{paper1} It is given by $F=F_0+F_D$,
where $F_0$ is the free energy in the absence of disorder and $F_D$ describes the coupling to
quenched defects. The disorder-free contribution is 
\begin{eqnarray}
\label{cleanFE}
F_0= \int_{\bf r}{\tilde{\epsilon}_1|{\bf t}({\bf r})|^2\over 2 n({\bf r})} 
+ {1\over 2 n_0^2}\int_{\bf q}{\Big\{} c_{44}^{c0}({\bf q}) |{\bf t}({\bf q})|^2 
+ c_{11}^0({\bf q})|\delta n ({\bf q})|^2 {\Big\}}\;,
\end{eqnarray}
where $\delta n({\bf r})=n({\bf r})-n_0$. The density and tilt field are related
by the familiar constraint that flux lines cannot start nor stop inside the sample,
\begin{equation}
\label{constraint}
\partial_z n({\bf r})+ {\bf \nabla}_{\perp}\cdot {\bf t}({\bf r})=0\;.
\end{equation}
The bare elastic constants, $c_{44}^{c0}$ (the compressive
part of the tilt modulus) and $c_{11}^0$ (the bare compressional modulus),
are determined
by the intervortex interaction. For isotropic materials they are simply 
$c_{44}^{c0}({\bf q})=c_{11}^{c0}({\bf q})=(B^2/4\pi)/(1+q^2\lambda^2)$, with $\lambda$
the screening length.\cite{notec11}
Quenched disorder from material defects couples to the flux-line density and 
gives a contribution
\begin{equation}
\label{disFE}
F_D= \int_{\bf r} V_D ({\bf r})\delta n ({\bf r}).
\end{equation}
The random potential $V_D({\bf r})$ is taken to be Gaussian, statistically homogeneous, and isotropic in the $xy$ plane so that 
\begin{equation}
\label{dicor}
\overline{\delta V_D({\bf q}) \delta V_D ({\bf q'})}=\Delta (q_{\perp}, q_z) \Omega \delta_{{\bf q}_{\perp}+{\bf q'}_{\perp}, 0}\;,
\end{equation}
where $\delta V_D ({\bf r})= V_D({\bf r})-\overline{V_D({\bf r})}$, and the
overbar represents the disorder average.
The correlator $\Delta(q_\perp,q_z)$ depends on the geometry of disorder and will be specified below for the case of interest.

The hydrodynamic free energy $F_0$ goes beyond the Gaussian hydrodynamic model
commonly used in the vortex literature \cite{MCM} as the first term on the right hand side of 
Eq. (\ref{cleanFE}), describing the 
``kinetic
energy'' part of the vortex interaction, incorporates non-Gaussian 
terms.
In a recent publication we showed that this non-Gaussian hydrodynamics
is precisely equivalent to the charged boson model of Feigel'man and coworkers.\cite{feigel}
Such a ``nonlinear hydrodynamics'' was used in Ref. \cite{paper1}
to evaluate perturbatively 
the enhancement of $c_{44}$ from interactions in a clean material. 
It was shown there that nonlocality is crucial to yield a correction to $c_{44}$ that remains finite for $L\rightarrow\infty$.
The same model is used here to evaluate corrections to $c_{44}$ from disorder.
In particular, for aligned columnar defects our calculation yields a criterion for 
locating the transition line $T_{BG}(B)$ {\it at all fields}.

The tilt modulus, $c_{44}$, can be expressed in terms of the tilt field autocorrelation function,
$T_{ij}({\bf q})=\overline{\langle t_i({\bf q}) t_j({\bf -q}) \rangle}$,
where the brackets denote a thermal average with weight $\sim\exp[-(F_0+F_D)/k_BT]$, 
to be carried out
subject to
the constraint (\ref{constraint}). 
It is given by
\begin{equation}
\label{longwavelc44}
{{n_0^2 k_B T}\over{c_{44}}}=\lim_{q_z \rightarrow 0} \lim_{q_{\perp} \rightarrow 0} 
P_{ij}^T(\hat{\bf q}_{\perp})T_{ij}({\bf q})\;,
\end{equation}
with $P_{ij}^T(\hat{\bf q}_{\perp})=\delta_{ij}-\hat{q}_{{\perp}i} \hat{q}_{{\perp}j}$
the familiar transverse projection operator, and ${\bf \hat{q}_\perp}={\bf q_\perp}/q_\perp$.
When only quadratic terms in the fluctuations $\delta n$ and ${\bf t}$ are retained,
the resulting Gaussian 
free energy, $F_{G}$, is 
$F_G=F_{0G}+F_D$,
where $F_{0G}$ is obtained from Eq. (\ref{cleanFE}) by the replacement 
$|{\bf t}|^2/n({\bf r}) \rightarrow |{\bf t}|^2/n_0$. 
The first two terms on the RHS of Eq. ({\ref{cleanFE}) can then be combined to define 
a bare tilt modulus $c_{44}^0=n\tilde{\epsilon}_1+c_{44}^{0c}$ and Eq. (\ref{longwavelc44})
is simply an identity.
To Gaussian order there is no coupling between the density field and 
the transverse part of the tilt field that determines $c_{44}$. As a result, disorder,
which couples to the density, does not change the tilt modulus. Non-Gaussian terms in the hydrodynamic free energy of Eq. (\ref{cleanFE})
do introduce a coupling between transverse tilt and density, yielding a renormalization
of the tilt modulus. Specifically, the free energy is written as 
$F=F_{G}+\delta F$, with
\begin{equation}
\label{deltaF}
\delta F=-\int_{\bf r}{\tilde{\epsilon}_1|{\bf t}({\bf r})|^2\over 2 n_0}
{\delta n({\bf r})\over n({\bf r})}. 
\end{equation}
By treating the non-Gaussian part $\delta F$ perturbatively, we obtain
\begin{equation}
{n_0^2\over c_{44}}={n_0^2\over c_{44}^0}\bigg[1-{\tilde{\epsilon}_1/n_0\over c_{44}^0/n_0^2}
{n_n\over n_0}\bigg],
\end{equation}
with 
\begin{equation}
\label{normalf}
{n_n\over n_0}={1\over n_0^2}\int_{\bf q}\bigg\{
\bigg[1-{\overline{\langle|{\bf t}({\bf q})|^2\rangle_G}\over 2 n_0\tilde{\epsilon}_1k_BT}\bigg]
\overline{\langle|\delta n({\bf q})|^2\rangle_G}
-{|\overline{\langle{\bf \hat{q}_\perp}\cdot{\bf t}({\bf q})\delta n(-{\bf q})\rangle_G}|^2\over 2n_0\tilde{\epsilon}_1k_BT}\bigg\}.
\end{equation}
The brackets $\overline{\langle ...\rangle_G}$ in Eq. (\ref{normalf}) denote a thermal average
with weight $\sim\exp[-F_G/k_BT]$, subject to the constraint (\ref{constraint}), 
followed by the average over quenched disorder. The correction has been denoted
by $n_n$ because it has the suggestive interpretation of a normal-fluid density.
Notice that its expression is formally identical to that obtained in \cite{paper1} in the absence
of disorder. Disorder only enters here in the expressions for the Gaussian
correlators, which can be found for instance in Ref. \cite{paper1}.
%
%\begin{equation}
%\label{dencor}
%\overline{\langle |\delta n({\bf q})|^2\rangle_G}=k_B T
%{{ n_0^2 q_{\perp}^2}\over{D_L({\bf q})}}+ \Delta ({\bf q})\bigg[{{n_0^2
%q_{\perp}^2}\over{D_L({\bf q})}}\bigg]^2\;,
%\end{equation}
%
%\begin{equation}
%\label{ttcor}
%\overline{\langle t_i({\bf q}) t_j({-\bf q})\rangle_G}= 
%{n_0^2k_B T \over{c_{44}^0({\bf q})}}P_{ij}^T({\bf q}_{\perp})
%+\bigg\{ {n_0^2k_B T q_z^2\over{D_L({\bf q})}} 
%+ \Delta({\bf q}){{q_{\perp}\over{q_z}}}\bigg[{{n_0^2 q_z^2}\over{D_L({\bf q})}}\bigg]^2
%\bigg\}P_{ij}^L({\bf q}_{\perp})\;,
%\end{equation}
%%
%\begin{equation}
%\label{dtncor}
%\overline{\langle {\bf \hat{q}_\perp}\cdot{\bf t} ({\bf q}) \delta n(-{\bf q})\rangle_G}=
%{ { n_0^2 k_BTq_{{\perp}}q_z}\over{D_L({\bf q})}}+ 
%\Delta({\bf q}){{q_{\perp}}\over{q_z}}\bigg[{{n_0^2 q_{{\perp}} q_z}\over{D_L({\bf q})}}\bigg]^2\;,
%\end{equation}
%
%where $1/D_L({\bf q})=1/[c_{44}^0({\bf q}) q_z^2+ c_{11}^0({\bf q}) q_{\perp}^2]$ is the longitudinal elastic ``propagator''. 
A more general expression for the renormalized tilt modulus at finite wave vector 
can also be obtained by the same methods.\cite{usunp}
In the dilute limit, $\lambda<<a$, with $a=1/\sqrt{n_0}$ the intervortex
spacing, the $z$-nonlocality becomes insignificant and Eq. (\ref{normalf})
reduces to the result obtained for instance by T\"auber and Nelson using the local boson mapping.\cite{TN}

We now focus on the case of correlated disorder created by heavy ion
irradiation yielding rectilinear damage tracks aligned with the field
${\bf H}_0$. This  corresponds to
$\Delta({\bf q})=\Delta L \delta_{q_z, 0}$,
where $\Delta \approx U_0^2 b^4/d^2$, with $U_0$ the depth of the pinning
potential well, $b$ its radius, and $d$ the average distance between
columnar defects.\cite{NV}
In this case 
the disorder contribution to $n_n$ from disorder
takes a particularly simple form, given by
\begin{equation}
\label{main}
{n_n^D\over n_0}=-{{\Delta}\over{2 \tilde{\epsilon}_1^2}}
\int_{\bf q_{\perp}}{{q_{\perp}^4}\over{[\epsilon_B(q_{\perp})/{k_B T}]^4}}G(q_{\perp})\;,
\end{equation}
where
$G(q_{\perp})=1+ c_{44}^{c0}(q_{\perp}, q_z=0)/  c_{44}^{0}(q_{\perp}, q_z=0)$,
and $\epsilon_B(q_{\perp})$ corresponds to the spectrum of $2d$ bosons with screened interactions,
\begin{equation}
\label{Bog}
{{\epsilon_B(q_{\perp})}\over{k_B T}}= 
\sqrt{{n_0q_\perp^2V(q_\perp)\over\tilde{\epsilon}_1}
+ \bigg({k_B Tq_{\perp}^2\over 2 \tilde{\epsilon}_1}\bigg)^2}\;,
\end{equation}
with
$V(q_{\perp})=V_0/(1 + \tilde{\lambda}_{\perp}^2 q_{\perp}^2)$
the screened boson interaction
and $V_0={\phi_0^2}/{4\pi}$.

The function $G(q_{\perp})$ depends only weakly on $q_{\perp}$, varying
between $1$ and $2$, and 
is the only manifestation of the $z$-nonlocality of
the intervortex interaction. If we let $G(q_\perp)=1$ and negelct the screening,
i.e., $V(q_\perp)=V_0$, our result becomes 
identical to the
disorder-induced renormalization of $1/c_{44}$ obtained by others.
\cite{Hwa,TN}
As expected, for the case of aligned
columnar defects the $z$-nonlocality is not very important.
The screening of the boson interaction incorporated in $V(q_\perp)$ is,
however, important at high vortex densities.
This becomes clear by plotting the boson spectrum as a function of wave
vector, as shown in Fig. 1. At high density the screening of the interaction 
yields a flat region in the spectrum. The wave vector integral in Eq.
(\ref{main}) is dominated by $q_\perp\sim k_{BZ}$. At low density the main contribution
to the integral comes therefore from a region where the spectrum is phonon-like,
i.e., $\epsilon_B(q_\perp)/(k_BT)\approx q_\perp\sqrt{n_0\tilde{V}(q_\perp,T)/\tilde{\epsilon}_1}$,
as obtained from the theory of uncharged boson superfluids.\cite{noteV0}
At high density, however, the main contribution to the integral comes from a 
region of wave vectors where the spectrum is plasmon-like, i.e.,
$\epsilon_B(q_\perp)/(k_BT)\approx \sqrt{4\pi n_0}$, as appropriate for a charged
superfluid. 
This was actually recognized earlier by Larkin and Vinokur,
who in order to use the results of the local boson theory at high fields
proposed an ad hoc formula for the Bogoliubov spectrum that 
captures the dense liquid physics by interpolating between these two limits.\cite{LV}
Hydrodynamics naturally provides a simple and unified description of the behavior of flux-line liquids in the
presence of disorder that applies over a wide range of
densities. 

As discussed in Ref. \cite{paper1}, the perturbation theory breaks down at high fields
and temperatures
($k_BT/\tilde{\epsilon}_1a>1$). An approximate expression for the tilt moduls of the form proposed
by Larkin and Vinokur \cite{LV} can be obtained by treating the nonlinearities in a mean
field approximation, as done by Feigel'man and collaborators for the charged boson
superfluid.\cite{feigel} The tilt-tilt autocorrelation function is evaluated by 
retaining only transverse fields and using a Hartree-type approximation for the non-Gaussian terms.\cite{usunp}
This yields
\begin{equation}
{n_0^2\over c_{44}}\approx{1\over c_{44}^{c0}/n_0^2+\tilde{\epsilon}_1/(n_0-n_n)},
\end{equation}
with $n_n$ given approximately by Eq. (\ref{normalf}).

The Bose glass transition line $B_{BG}(T)$ can now be evaluated
as the locus of points in the $(B,T)$ plane where $n_n^D/n_0=1$, 
corresponding to $c_{44}\rightarrow\infty$.
In general the integral in Eq. (\ref{main}) has to be evaluated numerically. The
resulting $B_{BG}(T)$ phase line is shown in Fig. 2. 
Analytical results can be obtained in the limit of low and high field.  
For a dilute liquid
$(\lambda>>a)$, the screening of the interaction can be
neglected and the Bogoliubov spectrum can be approximated as linear in $q_\perp$.
The integral can then be evaluated and gives
\begin{equation}
\label{BGdil}
B_{BG}(T)\approx B_\phi{\ln(1/n\lambda^2)\over 8\pi^2}\bigg({T^*\over T}\bigg)^4
\sim \bigg({T_c-T\over T}\bigg)^4,
\end{equation}
where $B_\phi=\phi_0/d^2$ is the matching field and 
$k_B T^*=b(\tilde{\epsilon}_1 U_0)^{1/2}$ is a characteristic pinning energy scale,
in agreement with earlier work by other authors.\cite{Hwa,LV,TN} 
In dense liquids ($\lambda>>a$), the Boguliobov spectrum 
is approximately $q_\perp$-independent.
Again, the transition line can be obtained analytically, with
\begin{equation}
\label{BGdense}
B_{BG}(T)\approx B_\phi {U_0b^2\over 64\pi\tilde{\epsilon}_1d^2}
    \bigg({T^*\over T}\bigg)^6\sim \bigg({T_c-T\over T}\bigg)^6,
\end{equation}
which agrees with the earlier result of  Larkin and Vinokur\cite{LV}.

Hydrodynamics provides a physically transparent framework for evaluating 
disorder-induced corrections
to the elastic constants of the vortex liquid. It naturally incorporates the full 
nonlocality of the intervortex interaction and it allows us to compute finite-wavevector
elastic constants for arbitrary disorder geometry. In particular, the nonlocality in the field
($z$) direction is expected to be important for splayed columnar defects, where terms 
coupling disorder directly to the tilt field may need to be incorporated in the 
coarse-grained theory. This work will be presented elsewhere.\cite{usunp}

\vspace{0.3in}
\noindent{\bf Figure Captions}
\begin{figure}[h]
\begin{center}\leavevmode
%\epsfxsize=6.5cm
%\epsffile{spectrum.ps}
%\psfig{file=spectrum.ps}
\end{center}
%\vspace{30mm}
\caption{The boson spectrum $\epsilon^*=\big[\epsilon_B(q_\perp)/(k_BT k_{BZ})\big]^2$ versus 
wavevector $\rm K=q_\perp/k_{BZ}$ for $n_0\lambda^2=0.1$
(solid line) and $n_0\lambda^2=10.$ (dashed line).}
\end{figure}
\begin{figure}[h]
\begin{center}\leavevmode
%\epsfxsize=6.5cm
%\epsffile{spectrum.ps}
%\psfig{file=spectrum.ps}
\end{center}
%\vspace{30mm}
\caption{The Bose glass transition line as obtained from the condition $n_n^D/n_0=1$.
The vertical axis is $B^*=B_{BG}(T)/B_\phi$ and the horizontal axis is
$t=T/T_c$.The parameter values used correspond to $B_{\phi}=2.3T$, $b=20{\AA}$,
and $T^*=0.7T_c$ at $T=0$. The transition line is well approximated by
$B_{BG}\sim 1/T^6$ as in Eq. (15) for all but extremely low fields.The
dashed line represents the the mean field $H_{c2}$. }
\end{figure}

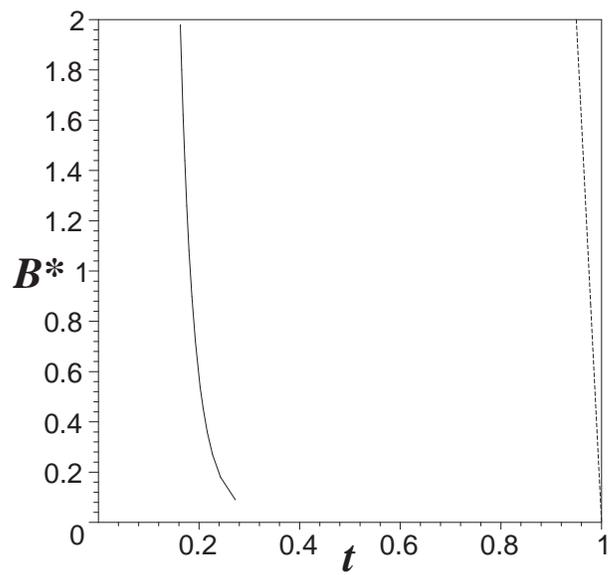
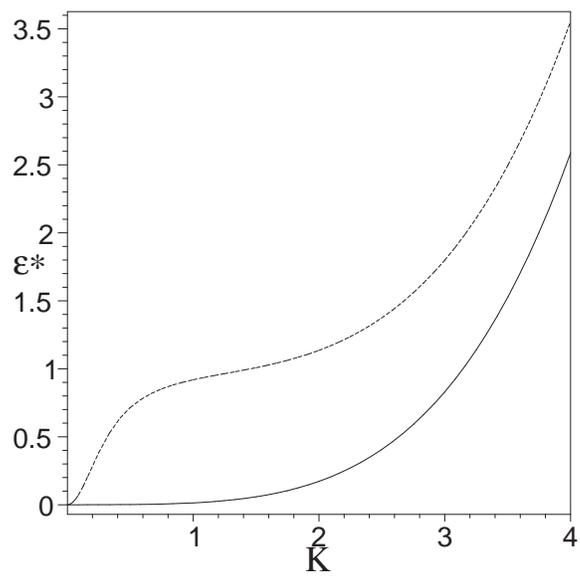
\end{document}